\documentclass[epj,nopacs]{svjour}

\usepackage{float}
\usepackage{amsmath}
\usepackage{amssymb}
\usepackage{cite}
\usepackage{graphicx}
\usepackage{multicol}
\usepackage[colorlinks=true,linkcolor=blue,citecolor=blue,urlcolor=blue]{hyperref}

\newcommand{\no}{\noindent}
\newcommand{\vsp}[1]{\vspace{#1}}
\newcommand{\hsp}[1]{\hspace{#1}}
\newcommand{\tx}[1]{\mathrm{#1}}

\begin{document}

\title{Application of Bayesian statistics to the sector of decay constants in three-flavour $\chi$PT}

\author{Mari\'an Koles\'ar\thanks{kolesar@ipnp.mff.cuni.cz} \and Jaroslav \v{R}\'{i}ha\thanks{jara.riha@email.cz}}

\institute{Institute of Particle and Nuclear Physics, Faculty of
Mathematics and Physics,\\ Charles University in Prague, V
Hole\v{s}ovi\v{c}k\'ach 2, 18000
Prague, Czech Republic.}

\abstract{
The sector of decay constants of the octet of light pseudoscalar mesons in the framework of 'resummed' $SU(3)$ chiral perturbation theory is investigated. A theoretical prediction for the decay constant of $\eta$-meson is compared to a range of available determinations. Compatibility of these determinations with the latest fits of the $SU(3)$ low energy coupling constants is discussed. Using a Bayesian statistical approach, constraints on the low energy coupling constants $L_4^r$ and $L_5^r$, as well as higher order remainders to the decay constants $F_K$ and $F_\eta$, are extracted from the most recent experimental and lattice QCD inputs for the values of the decay constants.}

\authorrunning{M. Kolesár \and J. Říha}
\titlerunning{Application of Bayesian statistics to the sector of decay constants in three-flavour $\chi$PT}
\maketitle

\section{Introduction}
Decay constants of the octet of light pseudoscalar mesons have a deep connection to the spontaneous breaking of chiral symmetry. Within the standard framework of \newline
$SU(3)_L\times SU(3)_R$ chiral perturbation theory ($\chi$PT) \cite{Gasser:1984gg}, the effective theory of quantum chromodynamics at low energies, the decay constants are directly connected to the renormalization and diagonalization of the kinetic part of the Lagrangian. 

Starting from the effective generating functional

\begin{align}
		e^{i Z_{eff}[v,a,s,p\,]}\ &=\ 
			\int\mathcal{D}U\ e^{i\int \mathrm{d}^4x\ \mathcal{L}_{eff}[U,v,a,s,p\,]},\\
		U(x)\ &=\ \tx{exp}{\left( \frac{i}{F_0}\,\phi^a(x)\lambda^a \right)},
\end{align}		
			
\no where $\phi^a(x)$ are the pseudo-Goldstone boson fields collected in a matrix field $U(x)$, one can obtain the connected $n$-point Green functions $G^{a_1 \dots a_n}_{P_1 \dots P_n}(p_1,\dots p_n)$ as on-shell \newline
residues of the Fourier transformed functional derivatives of $Z_{eff}[v,a,s,p\,]$ with respect to the axial vector sources $a_i$. $P_1 \dots P_n$ are the pseudo-Goldstone bosons in the in- and out-states with momenta $p_1,\dots p_n$. One then finds the following relation between the Green functions and the  the elements of the scattering matrix $A_{P_1 \dots P_n}(p_1,\dots p_n)$ \cite{Kolesar:2016jwe}

\begin{equation}
		G^{a_1 \dots a_n}_{P_1 \dots P_n}(p_1,\dots p_n) = F^{a_1}_{P_1} \dots 
		F^{a_n}_{P_n}A_{P_1 \dots P_n}(p_1,\dots p_n).
\end{equation}

\hfill\\ \\ \\[-1pt]

\no $F^{a_i}_{P_i}$ are the generalized decay constants which might in general include mixing terms. They correspond to the renormalization of the external legs of Feynman diagrams. Due to the mixing, it is necessary to distinguish between the on-shell particles in the in- and out-states $P_1 \dots P_n$ and the fields $\phi_{a_i}$, coupled to the axial-vector currents.   

The leading order (LO) effective Lagrangian has the form

\begin{equation}
\label{L2 Lagrangian}
		\mathcal{L}_{eff}^{(2)}\ =\ \frac{F^2_0}{4} \tx{Tr}[D_{\mu}U D^{\mu}U^+ + (U^+ \chi + \chi^+ U)].
\end{equation}
		
\no Here 

\begin{equation}
		\chi=2B_0\,\tx{diag}(m_u,m_d,m_s),
\end{equation}		
		
\no in the case when the scalar external sources are taken to be the quark masses. There is no mixing in the kinetic part of the effective Lagrangian at the leading order and thus it's straightforward to see that all the decay constants are equal to the low energy constant $F_0$ -- the fundamental order parameter of the broken chiral symmetry.  		
							
At the next-to-leading order (NLO), taking here the part containing the low-energy coupling constants (LECs) $L_1\dots L_{10}$

\newpage
											
\begin{equation}
\label{L4 Lagrangian}
	\begin{split}
	&\mathcal{L}_{eff}^{(4)}(L_1\dots L_{10})\ =\ L_1\,\tx{Tr}[D_{\mu}U^+ D^{\mu}U]^2\,+\\
	& + L_2\,\tx{Tr}[D_{\mu}U^+ D_{\nu}U]\,\tx{Tr}[D^{\mu}U^+ D^{\nu}U]\,+\\
	& + L_3\,\tx{Tr}[D_{\mu}U^+ D^{\mu}U D_{\nu}U^+ D^{\nu}U]\,+\\
	& + L_4\,\tx{Tr}[D_{\mu}U^+ D^{\mu}U]\,\tx{Tr}[\chi^+ U + \chi\,U^+]\,+\\
	& + L_5\,\tx{Tr}[D_{\mu}U^+ D^{\mu}U(\chi^+ U + U^+\chi)]\,+\\
	& + L_6\,\tx{Tr}[\chi^+ U + \chi\,U^+]^2\,+\\
	& + L_7\,\tx{Tr}[\chi^+ U - \chi\,U^+]^2\,+\\
	& + L_8\,\tx{Tr}[\chi^+ U\chi^+ U + \chi\,U^+\chi\,U^+]\,+\\
	& + iL_9\,\tx{Tr}[F^{\mu\nu}_R D_{\mu}U D_{\nu}U^+ + F^{\mu\nu}_L D_{\mu}U^+ D_{\nu}U]\,+\\
	& + L_{10}\,\tx{Tr}[U^+ F^{\mu\nu}_R U F^L_{\mu\nu}],
	\end{split}				
\end{equation}

\no $\pi^0$--$\eta$ mixing occurs in the kinetic part of the effective action as an isospin breaking effect, inversely proportional to the difference of light quark masses $m_d-m_u$. If the $\pi^0$--$\eta$ mixing is neglected, the only two terms contributing to the renormalization of the kinetic part are the ones proportional to the LECs $L_4$ and $L_5$.				

As can be seen, the sector of decay constants in the isospin limit can be viewed as the simplest self-contained subsystem of the theory, only involving two LECs at the leading order ($F_0$, $B_0$) and two at the next-to-leading order ($L_4$, $L_5$). Quite intriguingly, none of these four constants is known with very high certainty. As will be shown in more detail in what follows, at leading order a significant suppression of the order parameters, compared to the two-flavour values, is still possible or even probable, given the recent results from phenomenology \cite{Bijnens:2014lea} and lattice QCD \cite{FLAG:2021npn}. At next-to-leading order, the constant $L_4^r$ is expected to be small due to its suppression in the limit of large number of colours \cite{Ecker:1988te}, but if it is indeed the case is still unknown \cite{Bijnens:2014lea}. Depending on the above, the value of $L_5^r$ can also vary widely \cite{Bijnens:2014lea,FLAG:2019iem}.

The motivation of our work is to investigate this segment of $\chi$PT by Bayesian statistical methods, the question being how much can be told about the low-energy coupling constants just by restricting ourselves to this sector. We do not neglect the higher orders, but treat them as a source of statistical uncertainty, thus avoiding the large number of LECs appearing at next-to-next-leading order (NNLO) \cite{Amoros:1999dp}. The framework of 'resummed' $\chi$PT \cite{DescotesGenon:2003cg} is very well suited for such an approach.

Naturally, such a task might have been accomplished long ago, if the values of all the decay constants were known with sufficient precision. While that has indeed been the case for the pions and the kaons, the value of the decay constant of the $\eta$ meson was considered to be very uncertain due to its strong mixing with $\eta'$. In fact, in $\chi$PT calculations the $\eta$ decay constant has been usually treated using its chiral expansion, not as an independent observable \cite{Bijnens:2007pr,Kolesar:2016jwe}. Our crucial input is thus a recent calculation of the $\eta$--$\eta'$ sector on lattice QCD by the RQCD collaboration \cite{RQCD:2021qem}, which allows us to derive the $SU(3)$ decay constant $F_\eta$ with some confidence.

\newpage

This work is a continuation of our initial inquiries \cite{Kolesar:2008fu} and \cite{kolesar2019}, with the major new ingredients being the updated input for $F_\eta$ from lattice QCD and Bayesian statistical analysis, implemented in a numerical way. It can also be noted that the $\eta$ decay constant has not been used as input for the purpose of extraction of the low-energy parameters of the $SU(3)$ $\chi$PT until now.

The paper is organized in the following way -- Section \ref{sec:Resummed ChPT} provides a concise summary of our theoretical framework, while Section \ref{sec:Decay constants} introduces the sector of decay constants and connected phenomenology in a more detailed way. Our implementation of the Bayesian statistical analysis is outlined in Section \ref{sec:Bayes}, while the employed assumptions are discussed in Section \ref{sec:Assumptions}. Section \ref{sec:Results} then presents the results of the paper, which are subsequently summarized in Section \ref{sec:Summary}.

\section{Resummed $\chi$PT} \label{sec:Resummed ChPT}

We use an approach to chiral perturbation theory, dubbed 'resummed' $\chi$PT \cite{DescotesGenon:2003cg, DescotesGenon:2007ta, Kolesar:2016jwe}, which was proposed as a way to accommodate the possibility of an irregular convergence of the chiral expansion. Such a scenario might occur if some of the leading order LECs ($F_0$ or $B_0$) were suppressed to a sufficient degree, so that the leading order was not dominant in the chiral expansion. In such a case the chiral series should be handled carefully, as unexpectedly large higher orders might result from reordering of the expansion. In our case, we will assume a large range of possible values of the leading order constants, so various scenarios are naturally possible.

Let us summarized the procedure in a few points:

\begin{itemize}
    \item[$\bullet$] We use the standard $\chi$PT Lagrangian (\ref{L2 Lagrangian}, \ref{L4 Lagrangian}), based on the usual power counting $m_{q}\sim O(p^{2})$ \cite{Weinberg:1978kz}.
    \item[$\bullet$] Expansions of quantities related linearly to Green functions of QCD currents are trusted
    ("safe observables"). We assumed that for these expansions the NNLO and higher
    order terms are reasonably small, though not necessary negligible. Leading order terms are not required to be dominant.
    \item[$\bullet$] The expansions are expressed explicitly to next-to-leading order, all higher order contribution are summed into \textit{higher order remainders}. Thus for an observable $A$ the 'resummed' chiral expansion has the form

    \begin{equation}
        A = A^{(LO)}+A^{(NLO)}+A\delta A, \qquad \delta A \ll 1
    \end{equation}

    \item[$\bullet$] These higher order remainders will not be neglected, but estimated and treated as sources of error. In general, they might have a non-trivial analytical structure, though this is not the case for the decay constants. All higher order LECs are effectively contained in the remainders, the large number of NNLO constants is thus traded off for a relatively smaller number of remainders.
\end{itemize}

\newpage

\section{Decay constants \label{sec:Decay constants}}

Decay constants of the light pseudoscalar meson nonet, consisting of the pions, kaons, $\eta$ a $\eta'$, can be introduced in terms of the QCD axial-vector currents 

\begin{equation}
\label{axial currents}
		i p_{\mu} F_P^a = \langle\,0\,|\,A_{\mu}^a(0)\,|\,P,p\,\rangle,
\end{equation}

\no where $A_{\mu}^a=\bar{q}\gamma_\mu \gamma_5 \lambda^a q$. The pion and kaon decay constants take a straightforward form in the isospin limit and their values are very well established from either experimental data or lattice QCD calculations \cite{PDG2020,FLAG:2021npn}. In contrast, $\eta$ nad $\eta'$ decay constants were not very well known, until quite recently, due to significant mixing. A lot of theoretical and phenomenological work has thus been devoted to the $\eta$--$\eta'$ sector, see e.g. \cite{Leutwyler:1997yr,Feldmann:1998vh,Benayoun:1999au,Escribano:2005qq,Klopot:2012hd,Guo:2015xva,Escribano:2015yup,Bickert:2016fgy,Gu:2018swy}. This list, far from exhaustive, includes investigations of the sector in the $U(3)_L\times U(3)_R$ large $N_c$ framework as well as phenomenological studies, aiming to extract the values of the decay constants and related mixing angles from experimental inputs. The results of the phenomenological studies span quite a range of values, some of which are not compatible with others (see \cite{RQCD:2021qem} for a detailed overview).

Masses of the $\eta$ and $\eta'$ mesons in a scheme with a single mixing angle were obtained in lattice QCD simulations around a decade ago \cite{RBC-UKQCD:2010dd,Dudek:2011tt,UKQCD:2011sg}. However, until recently, to our knowledge only the EMT collaboration \cite{EMTC:2017bjt} has attempted to calculate the full sector of mixing parameters, which they did in the quark flavour basis. Finally, as already mentioned, a comprehensive study of the sector, which goes down to the physical pion mass, has now been published by the RQCD collaboration \cite{RQCD:2021qem}.

The $SU(3)$ decay constant $F_\eta$, which is the point of our interest, is defined identically to $F_\eta^8$ in (\ref{axial currents}) and can therefore be related to the mixing parameters in the $U(3)$ octet-singlet basis						
									
\begin{equation}
		F_\eta = F_{\eta}^8 = F_8\cos\vartheta_8.
\end{equation}

For the purpose of this work, our main input will be the recent lattice QCD determination of $F_\eta^8$ by the RQCD collaboration \cite{RQCD:2021qem}

\begin{equation}
		F_\eta^8 = (1.123 \pm 0.035)F_\pi \quad \mathrm{(RQCD21). \label{RQCD21}}
\end{equation}		

For comparison, we will also use two model dependent results from phenomenology
\begin{align}
\label{EGMS15} F_\eta^8 &= (1.18 \pm 0.02)F_\pi \quad \mathrm{(EGMS15)}\\
\label{EF05} F_\eta^8 &= (1.38 \pm 0.05)F_\pi \quad \mathrm{(EF05).}
\end{align}
\no Here, EGMS15 \cite{Escribano:2015yup} is a more recent determination which is representative of lower values of this observable, better compatible with (\ref{RQCD21}). On the other hand, EF05 \cite{Escribano:2005qq} lies on the opposite end of the spectrum and is an example of a very high value of $F_\eta^8$. Reported uncertainties are quite low in both cases and thus these results are essentially incompatible with each other. 
		
In the framework of $SU(3)_L\times SU(3)_R$ chiral perturbation theory \cite{Gasser:1984gg}, the chiral expansion of the pseudoscalar meson octet in the isospin limit can be written in the following way \cite{Kolesar:2008fu}
\begin{align}								
\label{Fpi} F_{\pi}^2 &= F_0^2( 1-4\mu_{\pi}-2\mu_K) \,+\\[5pt]
		 \nonumber &+ \, 8 m_\pi^2 \left( L_4^r(r+2)+L_5^r \right) + F_{\pi}^2\delta_{F_{\pi}}\\[10pt]
\label{FK} F_K^2 &= F_0^2\left(1-\frac{3}{2}\mu_{\pi}-3\mu_K-\frac{3}{2}\mu_{\eta}\right)\,+\\
		 \nonumber &+\, 8 m_\pi^2 \left(L_4^r(r+2)+\frac{1}{2}L_5^r(r+1)\right)+F_K^2\delta_{F_K}\\[10pt]
\label{Feta} F_{\eta}^2 &= F_0^2(1-6\mu_K)\,+\\
\nonumber &+\, 8 m_\pi^2 \left(L_4^r(r+2)+\frac{1}{3}L_5^r(2r+1)\right)+F_{\eta}^2\delta_{F_{\eta}}.			
\end{align}	
									
\no This form is obtained directly from the generation functional of two-point Green functions in the logic of 'resummed' approach to $\chi$PT \cite{DescotesGenon:2003cg}. A strict form of the chiral expansion is used, where the original parameters of the Lagrangian are retained, thus avoiding any reordering of the series. $F_P^2\delta_{F_P}$ are the sum of all higher orders, the higher order remainders, which are not neglected. These effectively contain all low-energy coupling constants at higher orders. It should be also noted that in this case the remainders are real constants with no analytical structure and no scale dependence.

Chiral logarithms are denoted as

\begin{equation}
		\mu_P = \frac{m_P^2}{32\pi^2F_0^2} \ln\left(\frac{m_P^2}{\mu^2}\right),
\end{equation}

\no where $m_P$ are the pseudoscalar masses at leading order. In particular 

\begin{align}
	\nonumber m_\pi^2 &= 2B_0\hat{m},\\
	m_K^2 &= B_0\hat{m}(r+1),\\
	\nonumber m_\eta^2 &= \frac{2}{3}B_0\hat{m}(2r+1),
\end{align}
\no with

\begin{equation}
	\hat{m} = \frac{m_u+m_d}{2},\quad
	r = \frac{m_s}{\hat{m}}.
\end{equation}
	
As can be seen from (\ref{Fpi}--\ref{Feta}), chiral expansions of the decay constants up to next-to-leading order does indeed depend only on the two leading-order and two next-to-leading order LECs -- $F_0$, $B_0$ and $L_4^r$, $L_5^r$, respectively. The sector of decay constants can thus be considered as a simple, self-contained system, which can be investigated on its own. 

For convenience, we introduce a reparametrization of the chiral order parameters $F_0$ and $B_0$

\begin{equation}
	\begin{split}
	X &= \frac{2\,\hat{m}F_0^2 B_0}{F_{\pi}^2M_{\pi}^2}\equiv\frac{2\,\hat{m}\Sigma_0}{F_{\pi}^2M_{\pi}^2},\\
	Z &= \frac{F_0^2}{F_\pi^2},\\
	Y &= \frac{X}{Z} = \frac{2B_0\hat{m}}{M_\pi^2} = \frac{m_\pi^2}{M_\pi^2},\label{XYZ}
	\end{split}	
\end{equation}

\no where $\Sigma_0$ is the three-flavour chiral condensate and $M_\pi$ is the physical pion mass. Such a reparametrization is convenient as the parameters $X$ and $Z$ are restricted to the range $(0,1)$. Furthermore, the so-called paramagnetic inequality \cite{DescotesGenon:1999uh} puts an upper bound in the form of the two-flavour LO LECs: 

\begin{equation}
\label{PIQ}
	\begin{split}
	Z &\equiv Z(3) < Z(2),\\
	X &\equiv X(3) < X(2),
	\end{split}	
\end{equation}

\no where the two-flavour parameters are defined analogously to the three-flavour ones in (\ref{XYZ}).

Standard approach to the chiral perturbation series usually assumes values of
$X$ and $Z$ reasonably close to one, with the leading order dominating the expansion. On the other hand, $Z=0$ would correspond to a restoration of chiral symmetry, while $X=0$ to a scenario with a vanishing chiral condensate, which also implies $Y=0$.

The most recent NNLO standard $\chi$PT fit \cite{Bijnens:2014lea} provides two different sets for the NLO LECs. It's based on a large number of inputs, including $\pi K$ and $\pi\pi$ scattering lengths, $K_{l4}$ form factors and pion scalar and vector form factors. It also uses the ratio $F_K/F_\pi$ (but not $F_\eta$). Overall, it uses 16 input observables to fit 8+34 NLO and NNLO parameters. The main fit (BE14) fixes $L_4^r$ by hand, in order to ensure the expected suppression in the large $N_c$ limit \cite{Ecker:1988te}. FF14 (free fit) releases this constraint. Their results for $L_4^r$ and $L_5^r$ are (at $\mu=770$ MeV):

\begin{equation}
\label{BE14}
	\begin{split} 
		10^3L_4^r &\equiv 0.3,\\
		10^3L_5^r &= 1.01\pm 0.06 \qquad \mathrm{(BE14)},
	\end{split}	
\end{equation}

and

\begin{equation}
\label{FF14}
	\begin{split}
		10^3L_4^r &= 0.76\pm 0.18,\\
		10^3L_5^r &= 0.50\pm 0.07 \qquad \mathrm{(FF14).}
	\end{split}	
\end{equation}

\no As can be seen, the obtained values are quite different. The difference is less pronounced for the LO LECs:

\begin{equation}
\label{BE14_LO}
	\begin{split}
		F_0 &= 71\ \mathrm{MeV},\\
		Y &= m_\pi^2/M_\pi^2 = 1.055 \qquad \mathrm{(BE14)}
	\end{split}	
\end{equation}
	
and	
	
\begin{equation}
\label{FF14_LO}
	\begin{split}
		F_0 &= 64\ \mathrm{MeV},\\
		Y &= m_\pi^2/M_\pi^2 = 0.937 \qquad \mathrm{(FF14).}
	\end{split}	
\end{equation}

For comparison, quite different values were obtained in \cite{Ecker:2013pba} by constructing $SU(3)$ amplitudes in a Large $N_c$ framework. $F_0$ was found to be very large ($88.1\pm4.1$ MeV) and $L_4^r$ compatible with zero ($(-0.05\pm0.22)\times 10^{-3}$).

The Flavour Lattice Averaging Group \cite{FLAG:2019iem, FLAG:2021npn} cites several lattice QCD determinations of $L_4^r$ and $L_5^r$. The last report \cite{FLAG:2021npn} highlights the results by HPQCD \cite{Dowdall:2013rya}:

\begin{equation}
\label{HPQCD 13A}
	\begin{split} 
		10^3L_4^r &= 0.09\pm 0.34,\\
		10^3L_5^r &= 1.19\pm 0.25 \qquad \mathrm{(HPQCD\ 13A)}
	\end{split}	
\end{equation}

and MILC \cite{MILC2010} 

\begin{equation}
\label{MILC 10}
	\begin{split}
		10^3L_4^r &= -0.02\pm 0.56,\\
		10^3L_5^r &= 0.95\pm 0.41 \qquad \mathrm{(MILC\ 10).}
	\end{split}	
\end{equation}

The leading-order LECs have also been recently calculated on lattice by the $\chi$QCD collaboration \cite{CHQCD:2021pql}. Though the results have not been fully published yet, the work has been cited by the Flavour Lattice Averaging Group \cite{FLAG:2021npn} with a favorable rating. While there are several other older determinations, for example by the MILC collaboration \cite{MILC2009,MILC2009A,MILC2010} or based on RBC/UKQCD \cite{Bernard:2012fw}, $\chi$QCD provides the first highly-rated calculation of these parameters in more than a decade, as far as we are aware of. The results were quoted by FLAG in the following form: 
		
\begin{equation}
\label{ChQCD}
	\begin{split}	
	F_0 &= 67.8(1.2)(3.2)\ \mathrm{MeV},\\
	\Sigma_0^{1/3} &= (F_0^2 B_0)^{1/3} = 232.6(0.9)(2.7)\ \mathrm{MeV}. 
	\end{split}
\end{equation}
																
\no We will use these values as alternative inputs for the \newline leading-order LECs.		
																
The purpose of this work is twofold - first, we will show that the 'resummed' $\chi$PT framework leads to a simple, but robust prediction for $F_\eta$. Then we will use the values of $F_\eta^8$ (\ref{RQCD21}--\ref{EF05}) as an input and use Bayesian statistical inference to obtain constraints on the higher order remainders $\delta_{F_K},\delta_{F_{\eta}}$ and the NLO LECs $L_4^r$ and $L_5^r$. We will compare these results with the two versions of the fit \cite{Bijnens:2014lea} (BE14 and FF14) and lattice QCD values (HPQCD 13A \cite{Dowdall:2013rya} and MILC 10 \cite{MILC2010})  and thus check the compatibility of the various values of $F_\eta$ and NLO LECs.

\section{Bayesian statistical analysis \label{sec:Bayes}}	

We use a statistical approach based on the Bayes' theorem \cite{DescotesGenon:2003cg,Kolesar:2017xrl}																							
\begin{equation}
\label{Bayes_theorem}
		P(X_i|\mathrm{data}) = \frac{P(\mathrm{data}|X_i)P(X_i)}{\int \mathrm{d}X_i\,P(\mathrm{data}|X_i)P(X_i)},
\end{equation}		
		
\no where $P(X_i|\mathrm{data})$ is the probability density function (PDF) of an explored set of theoretical parameters $X_i$ having a specific value given some experimental data. 

In the case of independent experimental inputs, \newline $P(\mathrm{data}|X_i)$ is the known probability density of obtaining the observed values of the observables $O_k$ in a set of experiments with uncertainties $\sigma_k$ under the assumption that the true values of $X_i$ are known, typically given as a normal distribution																		
\begin{equation}
	P(\mathrm{data}|X_i) = \prod_k\frac{1}{\sigma_k\sqrt{2\pi}}\,
	\mathrm{exp}\left[-\frac{(O_k^\mathrm{exp}-O^\mathrm{th}_k(X_i))^2}{2\sigma_k^2}\right].	
\end{equation}
		
$P(X_i)$ in (\ref{Bayes_theorem}) are prior probability distributions of $X_i$. We use them to implement theoretical assumptions, available experimental information and uncertainties connected with our parameters.

Traditionally, the prior has been understood as a degree of subjective belief. However, in our view, one does not necessarily needs to 'believe' in the validity of the prior in a scientific context, which we think can then be more appropriately interpreted as the quantification of available information and beyond that, the assumptions entering the analysis. Naturally, predictions might depend on the assumptions used. The Bayesian formalism allows us to straightforwardly implement a variety of assumptions and explore their consequences, which we consider to be an important feature of this approach. 

In our case we have three observables in the form of the three decay constants $F_\pi$, $F_K$, $F_\eta$. We will consider the ratios of quark masses as known and thus we are left with the following free theoretical parameters:

\begin{itemize}
	\item[--] leading order: $Z$, $Y$
	\item[--] next-to-leading order: $L_4^r$, $L_5^r$
	\item[--] higher orders: $\delta_{F_\pi}$, $\delta_{F_K}$, $\delta_{F_\eta}$.
\end{itemize}

\no As discussed in the next section, we will use several assumptions about these parameters, which will determine the prior distributions. 

Our implementation of the Bayesian statistical analysis is numerical. It consists of two steps - first we numerically generate a large ensemble of theoretical predictions $O^\mathrm{th}_k(X_i)$ for the decay constants, depending on the free parameters, and then we calculate the probability density functions (\ref{Bayes_theorem}), effectively using Monte Carlo integration.

\section{Assumptions \label{sec:Assumptions}}		

For the LO LECs $F_0$ and $B_0$, we use similar theoretical constraints as in \cite{Kolesar:2017xrl}, which define our priors for these parameters. Their approximate range then is  
															
\begin{align}
	&0 < Y < Y_{\mathrm{max}} \simeq 2.5, \label{Range_Y} \\
	&0 < Z < Z(2) = 0.86 \pm 0.01,\label{Range_Z}	\\
	&0 < X < X(2) = 0.89 \pm 0.01,\label{Range_X}
\end{align}		

\no where the explicit form of $Y_{\mathrm{max}}$, derived in \cite{DescotesGenon:2003cg}, is

\begin{equation}
		Y_\mathrm{max} = \frac{8F_K^2 M_K^2\left(\delta_{M_K}-1\right)-2F_\pi^2 M_\pi^2\left(r+1\right)^2\left(			\delta_{M_\pi}-1\right)}{M_\pi^2\left(r+1\right)\left(2F_K^2\left(\delta_{F_K}-1\right)-F_\pi^2(r+1)		\left(\delta_{F_\pi}-1\right)\right)}.
\label{Y_max_Descotes}
\end{equation}

\no Here $\delta_{M_\pi}$ and $\delta_{M_K}$ are higher order remainders for the chiral expansions of the pseudoscalar masses, which we treat analogously to the remainders of the decay constants (see (\ref{delta}) below).

In order to calculate the prior distributions, we consider $X$ and $Z$ as the primary variables:

\begin{equation}
	\begin{split}
		&P(Y,Z|\mathrm{data})~\mathrm{d}Y\mathrm{d}Z \equiv\\
		&\equiv P(X,Z|\mathrm{data})|_{X\to ZY}~\mathrm{d}Y Z\mathrm{d}Z = \\
		&= P(X,Z|\mathrm{data})~\mathrm{d}X \mathrm{d}Z.
	\end{split}
\label{P(Y,Z)}
\end{equation}
							
\begin{figure}
    \begin{center}
    \includegraphics[scale=0.6]{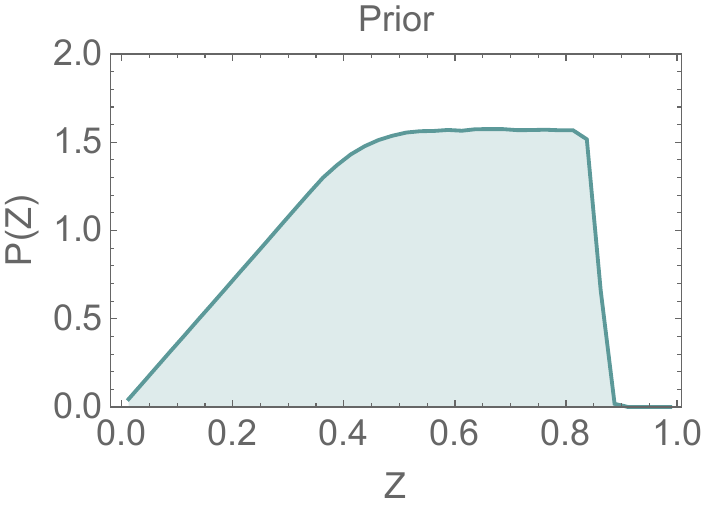} \\[5pt]
    \includegraphics[scale=0.6]{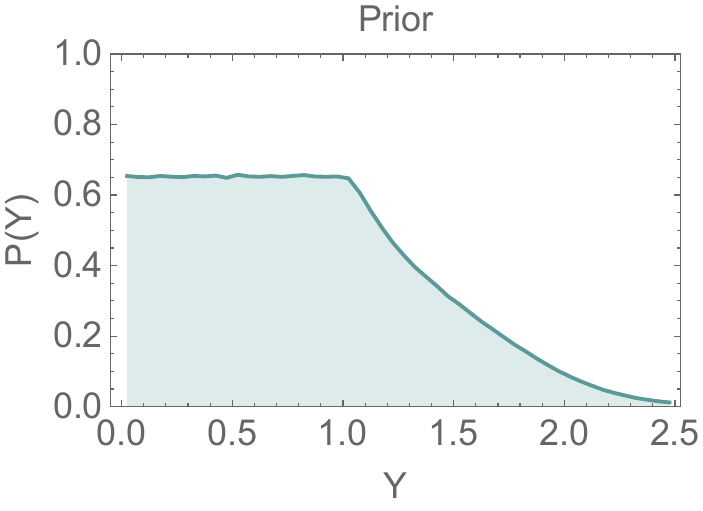}
    \end{center}
    \caption{Prior distributions based on (\ref{Range_Y}--\ref{P(Y,Z)}).}
\label{Fig_priors_PIQ}
\end{figure}

One might naturally ask, why not use $Y$ and $Z$ as the primary variables, as these are the free parameters in our case. That would mean using uniform distributions in the range $0 < Y < Y_{\mathrm{max}}$ and $0 < Z < 1$ as the a priori assumption. However, this leads to a quickly rising probability distribution for $X=ZY$ towards zero. Quite clearly, a very small chiral condensate is not a reasonable initial expectation. On the other hand, starting with uniform distributions for $0 < X < 1$, $0 < Z < 1$ and adding $Y < Y_{\mathrm{max}}$	ensures a relatively flat prior for $X$ and a vanishing distribution for $Z$ at $Z=0$. That we find reasonable, as it excludes the scenario with unbroken chiral symmetry and thus a world without the pseudo-Goldstone bosons. Then by including the paramagnetic inequality (\ref{PIQ}) we obtain the set (\ref{Range_Y}--\ref{Range_X}). These assumptions lead to probability distributions for the priors depicted in Figure \ref{Fig_priors_PIQ}.

For the purpose of obtaining constraints on the NLO LECs $L_4^r$ and $L_5^r$, we will use the determination of $Y$ from $\eta\to 3\pi$ decays \cite{Kolesar:2017xrl} as an additional assumption

\begin{equation}
		Y\ =\ 1.44 \pm 0.32\qquad\qquad (\eta\to3\pi). \label{Y_eta3pi}
\end{equation}		

\begin{figure}
    \begin{center}
    \hsp{-0.5cm} \includegraphics[scale=0.6]{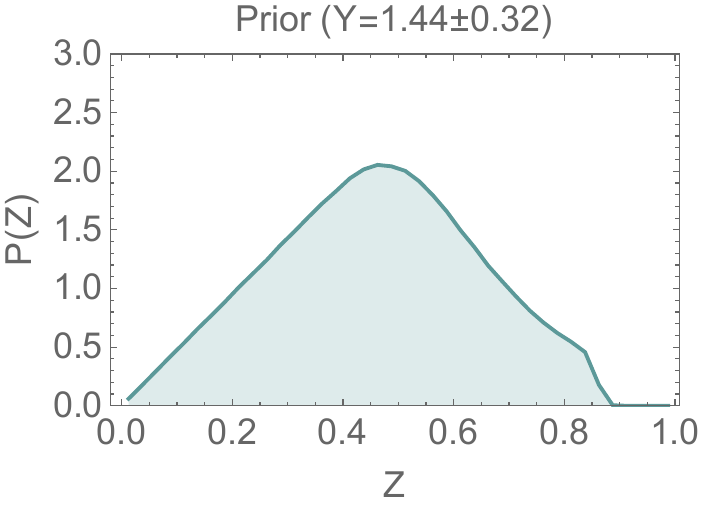} \\[5pt]
    \hsp{-0.5cm} \includegraphics[scale=0.6]{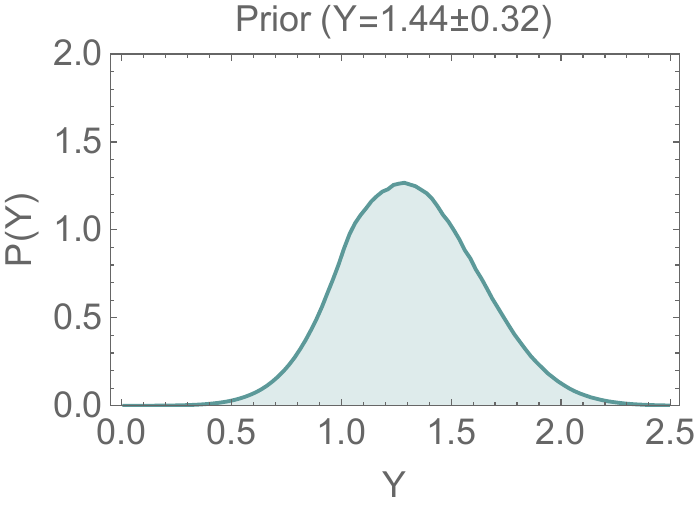}		\end{center}
    \caption{Prior distributions based on (\ref{Range_Y}--\ref{P(Y,Z)}, \ref{Y_eta3pi}).}
\label{Fig_priors_Y_eta3pi}
\end{figure}

\no This value was obtained from a Bayesian analysis of the $\eta\to 3\pi$ decay widths of the two decay channels and the Dalitz parameter $a$ in the charged channel. The tendency towards $Y>1$ is ultimately tied to the very large overall experimental decay rate compared to the simple estimate at leading order, given that the isospin violating parameter 
 $R$ is now known with a fairly good precision from lattice QCD \cite{FLAG:2021npn}. However, while this value is higher than the result of the fits BE14 and FF14 (\ref{BE14_LO}--\ref{FF14_LO}). it is compatible and does provide us with a reasonable uncertainty range. As can be seen in Figure \ref{Fig_priors_Y_eta3pi}, this input effectively excludes very low values of $Y$, which correspond to a significantly suppressed chiral condensate. In other words, such a scenario can be understood to be excluded by the phenomenology of the $\eta\to 3\pi$ decays.

As an alternative, we will also use the recent result by the $\chi$QCD Collaboration (\ref{ChQCD}), expressed in the form:

\begin{equation}
	\label{ChQCD_ZY}
	\begin{split}
		Y &= 0.95 \pm 0.10,\\
		Z &= 0.54 \pm 0.05 \qquad\mathrm{(\chi QCD21)}. 
	\end{split}
\end{equation}

\begin{figure}
    \begin{center}
    \includegraphics[scale=0.6]{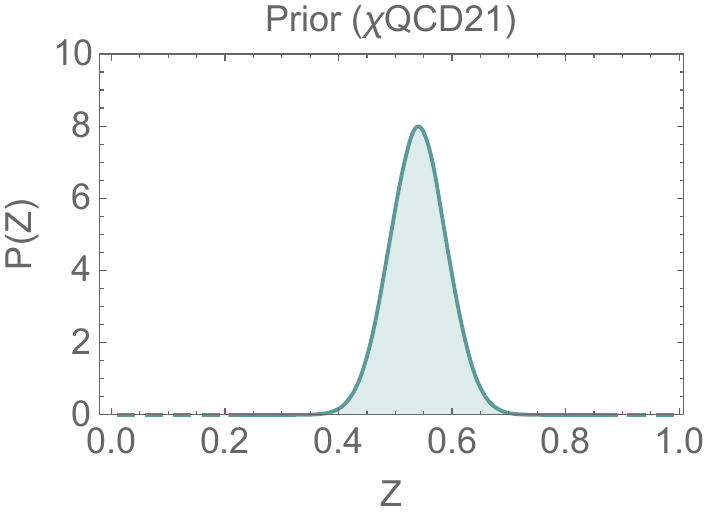}\\[5pt]
    \hsp{0.15cm}	
    \includegraphics[scale=0.6]{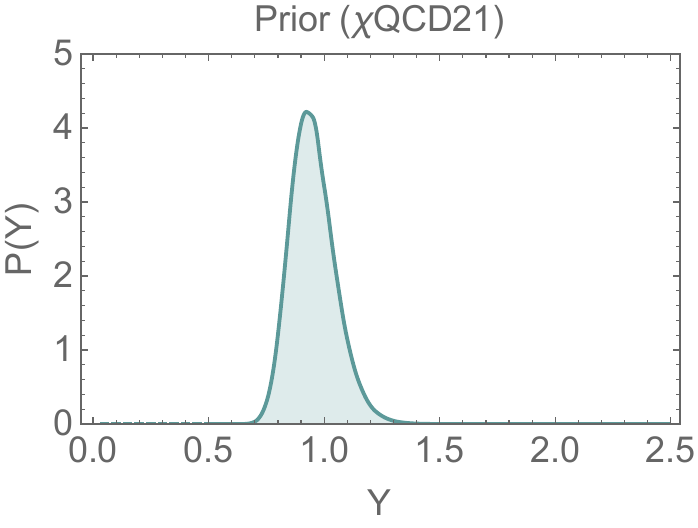}	
    \end{center}
    \caption{Prior distributions based on (\ref{Range_Y}--\ref{P(Y,Z)}, \ref{ChQCD_ZY}) ($\chi$QCD21).}
\label{Fig_priors_CHQCD}
\end{figure}

\no In comparison with our main input (\ref{Y_eta3pi}), $\chi$QCD21 also excludes high values of $Y$ and fixes $Z$ in a narrow range at a fairly low value. The prior distributions can be seen in Figure \ref{Fig_priors_CHQCD}. 

The three sets of priors introduced above should be understood as proceeding from more conservative to more restricted. The set (\ref{Range_Y}--\ref{P(Y,Z)}), is based on very general consideration rooted in QCD. The additional inputs (\ref{Y_eta3pi}) or alternatively (\ref{ChQCD_ZY}) then supplement an assumption about the value of the leading order LECs. Here, (\ref{Y_eta3pi}) is a more conservative one, effectively only excluding very low values of Y, based on results from $\eta\to 3\pi$ decays. It is compatible with all values quoted in Section \ref{sec:Decay constants}. On the other hand, (\ref{ChQCD_ZY}) uses very specific values from a recent lattice QCD study, which might be in tension with some other determinations. Our main results will be based on the more conservative assumption of the first two sets of priors, while the third one will be used to explore the consequence of assuming a particular value of the chiral order parameters, which are not yet firmly established, though.

Also, it should be noted that while the PDFs depicted on Figures \ref{Fig_priors_PIQ}-\ref{Fig_priors_CHQCD} illustrate the form of the priors for the parameters, they are not the actual inputs. The priors are the probabilistic conditions (\ref{Range_Y}-\ref{ChQCD_ZY}), from which the PDFs follow, but these do not capture the full information encoded in the multidimensional conditions (\ref{Range_Y}-\ref{ChQCD_ZY}).

Furthermore, while the Bayes theorem  (\ref{Bayes_theorem}) separates the probability distributions $P(\mathrm{data}|X_i)$, which depend on the theoretical predictions $O^\mathrm{th}_k(X_i)$, and the priors $P(X_i)$, it's quite desirable to implement the priors on the level of theoretical predictions, thus effectively incorporate the priors $P(X_i)$ into $P(\mathrm{data}|X_i)$, so one can examine the theoretical predictions including a realistic set of assumptions. Thus we implement the relations (\ref{Range_Y}-\ref{P(Y,Z)}), i.e. our default prior, when numerically generating the theoretical predictions $O^\mathrm{th}_k(X_i)$. 

As for the NLO LECs $L_4^r$ and $L_5^r$, where our goal is to extract constraints, we limit them to the range (at $\mu=770$~MeV)

\begin{align}
    10^{3} L_5^r &\in (0,2),\\
    10^{3} L_4^r &\in (-0.5, 2),
\end{align}

\no which we implement as a uniform distribution. The choice of the uniform distribution signifies the lack of preference for any particular value in the allowed range, which we think is appropriate given the range of values for these parameters available in the literature. Hence we do not use any particular value as our prior and the results are therefore not directly dependent on previous analyses.

The allowed range was chosen in a way to cover the values of all recent determinations we are aware of, including NNLO $\chi$PT \cite{Bijnens:2014lea} and lattice QCD \cite{FLAG:2019iem,FLAG:2021npn}. The upper bound is high enough to contain the PDFs of our main results. The distribution is cut off in particular cases where a lower bound for $L_5^r$ is obtained (see later), but in such instances we find reasonable to stick with a more conservative result.

We find no credible reason to assume $L_5^r$ smaller than zero. Our basic assumption is that $L_5^r$ is positive, which is consistent with available determinations, which are all clearly larger than zero.

The situation is more subtle concerning $L_4^r$. As commented above, this constant is suppressed in the large $N_c$ limit and some results indeed find its value close to zero or even slightly negative \cite{FLAG:2021npn, Ecker:2013pba}. From the paramagnetic inequalities (\ref{PIQ}) it follows  that there is a critical value of $L_4^r$ \cite{Descotes-Genon:2002nkp, DescotesGenon:2003cg}. For the value of $r$ we use (see (\ref{r}) below) we find:

\begin{equation}
		{L_4^r}^{(crit)} = -0.50\times 10^{-3}.
\end{equation}	

\no Hence we use a lower bound $L_4^r>-0.5\times 10^{-3}$.

We estimate the higher order remainders statistically, based on general arguments about the convergence 
of the chiral series \cite{DescotesGenon:2003cg}. Our initial ansatz is

\begin{equation}
	\delta_{F_P} = 0.0\pm 0.1. \label{delta}
\end{equation}		

\no We implement this by normal distributions, therefore the remainders are limited only statistically, not by any upper bound. However, our initial analysis will provide us with constraints on the higher order remainders obtained from the data for the decay constants. Subsequently, we will reuse these constraints as priors for the determination of NLO LECs $L_4^r$ and $L_5^r$, thus effectively shifting the initial ansatz (\ref{delta}).

We use the lattice QCD average \cite{FLAG:2021npn} for the value of the strange-to-light quark mass ratio $r$ 

\begin{equation}
\label{r}
		r= 27.23 \pm 0.10.
\end{equation}

Finally, the inputs for the pion and kaon decay constants are \cite{PDG2020}

\begin{equation}
\label{F_P_data}
	\begin{split}
		F_\pi &= 92.32 \pm 0.09\ \mathrm{MeV},\\
		F_K &= 110.10 \pm 0.21\ \mathrm{MeV}.
	\end{split}
\end{equation}

\no We use inputs from PDG \cite{PDG2020} for the masses of the particles as well, with the experimental uncertainties being negligible compared to other sources of error.

\section{Results \label{sec:Results}}

\subsection{Prediction for \boldmath{$F_\eta$}}

We will employ several ways of dealing with the system of equations (\ref{Fpi}--\ref{Feta}). At the first stage, it is possible to eliminate $F_0$, $L_4^r$ and $L_5^r$ by simple algebraic manipulations and thus we obtain a single equation
		
\begin{equation}
\label{Feta_eq}
	\begin{split}		 
		 F_{\eta}^2 &= \frac{1}{3}\Big[4F_K^2-F_{\pi}^2 + \frac{M_{\pi}^2 Y}{16\pi^2}
		\left(\ln \frac{m_{\pi}^2}{m_K^2}\,+\,(2r+1)\ln \frac{m_{\eta}^2}{m_K^2}\right) +\\
		 &+\,3F_\eta^2\delta_{F_{\eta}} - 4F_K^2\delta_{F_K} + F_\pi^2\delta_{F_{\pi}}\Big]. 
	\end{split}
\end{equation}	
																
\begin{figure}[b]
    \begin{center}													\includegraphics[scale=0.75]{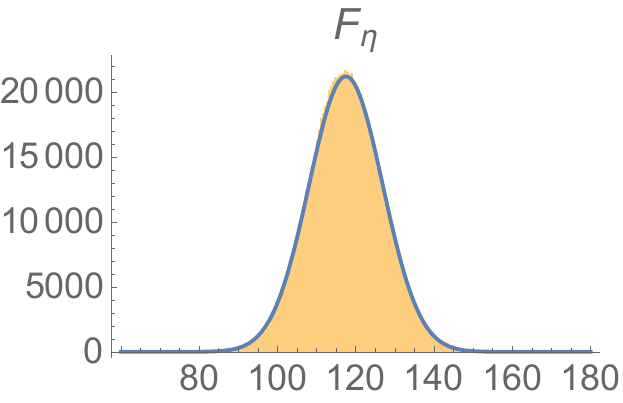}							\end{center}
    \caption{Theoretical prediction for $F_\eta$ ($10^6$ points). Gaussian fit overlaid.}
\label{Fig_Feta}	
\end{figure}

\no The equation depends, beyond the remainders $\delta_{F_P}$, only on a single parameter $Y$ and the dependence is very weak, as already noted in \cite{DescotesGenon:2003cg} and \cite{Kolesar:2008fu}. A histogram of $10^6$ numerically generated theoretical predictions is depicted in Figure \ref{Fig_Feta}, where the default assumptions (\ref{Range_Y}--\ref{P(Y,Z)}) (illustrated in Fig. \ref{Fig_priors_PIQ}) and (\ref{delta}--\ref{F_P_data}) were used. A Gaussian fit leads to a value 

\begin{equation}
		F_\eta = 117.5\pm 9.4\ \mathrm{MeV} = (1.28\pm0.10)F_\pi.
\end{equation}

\no This is an improved prediction over \cite{Kolesar:2008fu} and lies in between the values of EGMS15 (\ref{EGMS15}) and EF05 (\ref{EF05}), discussed above, while still being compatible with RQCD21 (\ref{RQCD21}). 

As noted, this result depends only very weakly on the value of $Y$ and thus the choice of the prior. E.g., adding the most restricting assumption (\ref{ChQCD_ZY}), illustrated in Fig. \ref{Fig_priors_CHQCD}, leads to an almost identical prediction

\begin{equation}
		F_\eta = 117.7\pm 9.3\ \mathrm{MeV} \qquad \mathrm{(\chi QCD21).}
\end{equation}

\subsection{Higher order remainders}

\begin{figure}
    \begin{center} 
    \vsp{0.5cm}
    \includegraphics[scale=0.5]{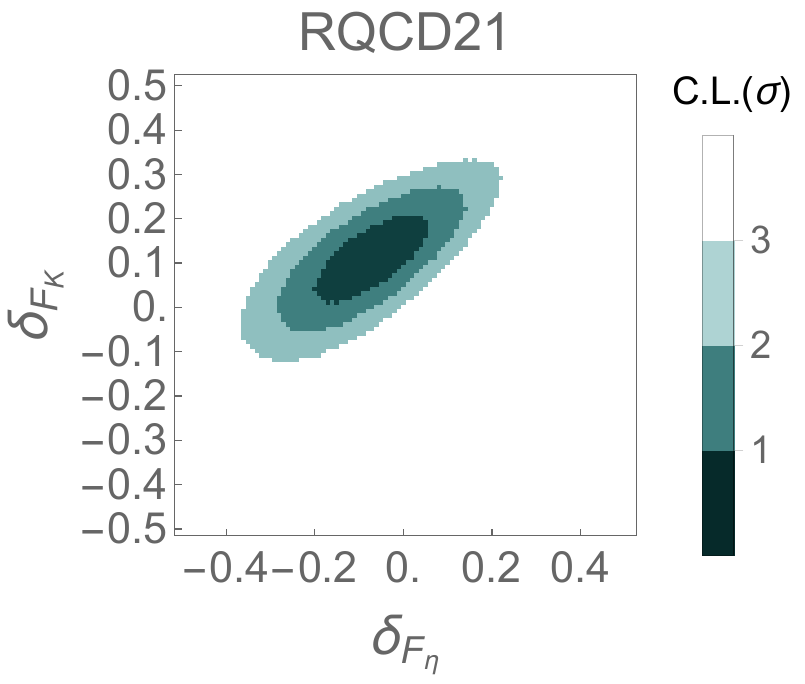}	\\[10pt]	\includegraphics[scale=0.5]{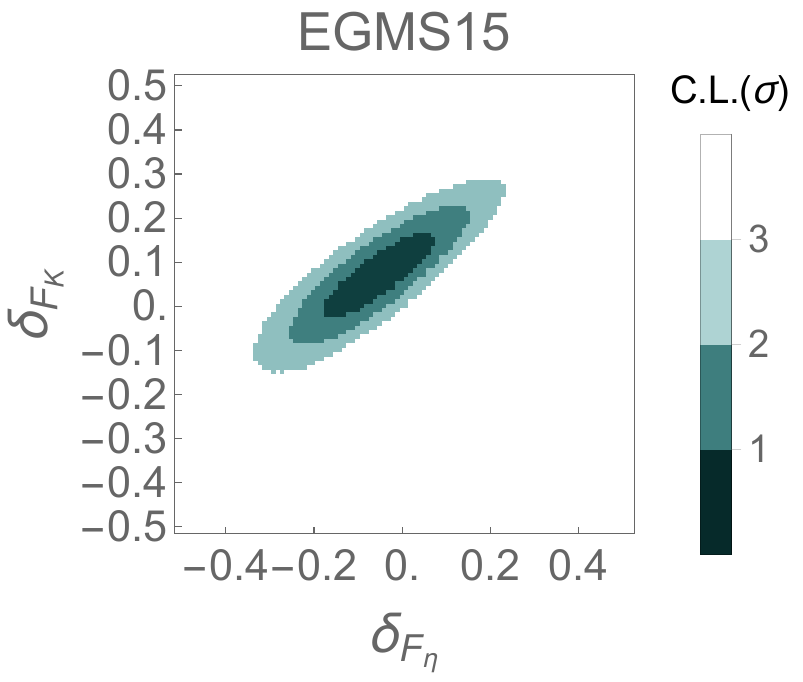}	\\[10pt]
    \includegraphics[scale=0.5]{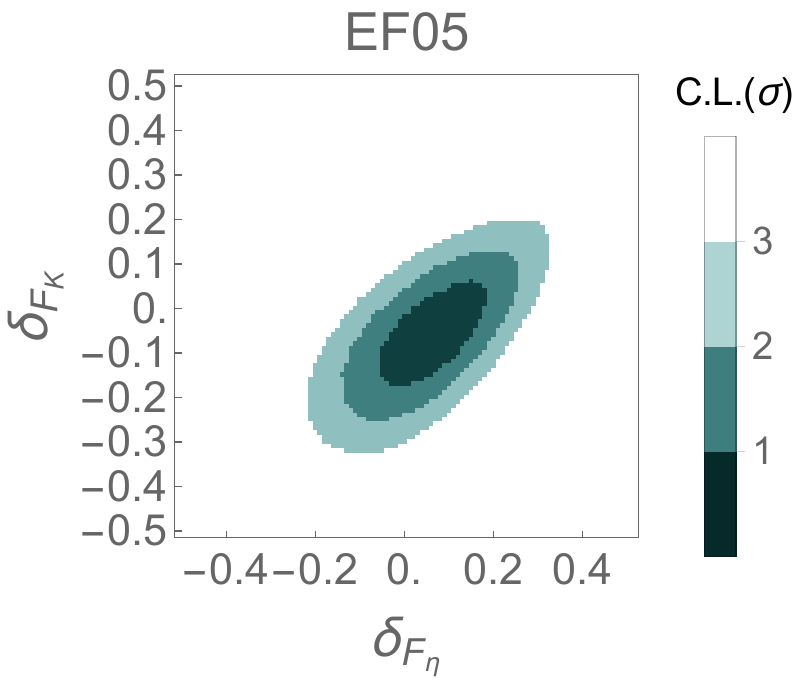}						\end{center}
    \caption{Constraints on higher order remainders from (\ref{Feta_eq}), alternative inputs for $F_\eta$.}
\label{Fig_delta}
\end{figure}

Next, given the weak dependence of (\ref{Feta_eq}) on $Y$, we can use RQCD21 (\ref{RQCD21}), EGMS15 (\ref{EGMS15}) and EF05 (\ref{EF05}) as alternative inputs for $F_\eta$ and employ the Bayesian statistical approach to extract information about the remainders. A contour plot with confidence levels can be found in Figure \ref{Fig_delta}, which leads to
\begin{align}
	\nonumber \delta_{F_K} &= 0.10 \pm 0.07,\\
	\label{delta_RQCD21}\delta_{F_\eta} &= -0.08 \pm 0.08,\\
	\nonumber \rho &= 0.71 \hsp{2cm} \mathrm{(RQCD21)},\\[10pt]
	\nonumber \delta_{F_K} &= 0.07 \pm 0.06,\\ 
	\label{delta_EGMS15} \delta_{F_\eta} &= -0.06 \pm 0.08,\\
	\nonumber \rho &= 0.85 \hsp{2cm} \mathrm{(EGMS15)},\\[10pt]
	\nonumber \delta_{F_K} &= -0.06 \pm 0.08,\\  
	\label{delta_EF05} \delta_{F_\eta} &= 0.05 \pm 0.08,\\
	\nonumber \rho &= 0.64 \hsp{2cm} \mathrm{(EF05),}
\end{align}	

\no where $\rho$ is the correlation coefficient. These values are compatible with the prior assumption (\ref{delta}). We can also compare these results with the NNLO contributions for $F_K$ obtained in \cite{Bijnens:2014lea}

\begin{align}
	F_K/F_\pi &= 1 + 0.176 + 0.023\quad \mathrm{(BE14)},\\
	F_K/F_\pi &= 1 + 0.121 + 0.077\quad \mathrm{(FF14)}.
\end{align}	

\no As can be seen, both are positive, while EF05 (\ref{delta_EF05}) implies a negative remainder $\delta_{F_K}$. It should be noted, however, that the work \cite{Bijnens:2014lea} uses a different form of the chiral expansion and thus this can only be taken as an indication that lower values of $F_\eta$ might be better compatible with the fits BE14/FF14.

In the following, we will use the relation (\ref{Feta_eq}) as an additional constraint, thus effectively implementing the results (\ref{delta_RQCD21}--\ref{delta_EF05}). In this way we obtain improved priors for the higher order remainders, compared to the initial ansatz (\ref{delta}).

\subsection{Extraction of \boldmath{$L_5^r$}}	

\begin{figure}
    \begin{center}		
    \includegraphics[scale=0.85]{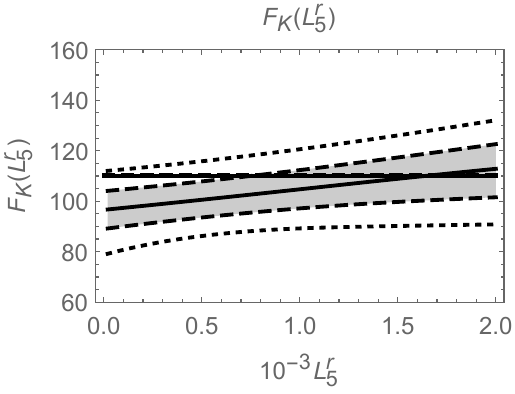} \\[10pt]
    \includegraphics[scale=0.85]{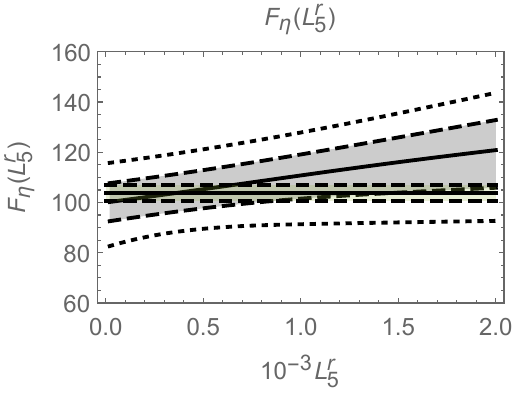}
    \end{center}
    \caption{Theoretical predictions for $F_K$ and $F_\eta$ with 1$\sigma$ CL (shaded+dashed) and 2$\sigma$ CL (dotted) contours depicted.\\ Horizontal lines -  data from \cite{PDG2020,RQCD:2021qem}.}
\label{Fig_L5_predictions}
\end{figure}

As a second step, we can algebraically eliminate $F_0$ and $L_4^r$ by using equation (\ref{Fpi}), which leads to a system of two equations for $F_K$ and $F_\eta$, now depending on $Y$, $L_5^r(\mu)$ and the remainders $\delta_{F_P}$. We numerically generated $10^8$ theoretical predictions for the kaon and eta decay constants (at $\mu=770$ MeV), shown in Figure \ref{Fig_L5_predictions}, in comparison with the data (\ref{RQCD21}) and (\ref{F_P_data}). Once again, the default priors (\ref{Range_Y}-\ref{P(Y,Z)}) has been implemented here, along with the assumptions  (\ref{delta}--\ref{F_P_data}). 

Our first task is to verify the general compatibility of our ensemble of theoretical predictions with the data. As can be seen in Figure \ref{Fig_L5_predictions}, this is indeed quite clearly the case, as  the predictions are compatible with the data in the whole range
of values. The data for $F_K$ are a little better compatible with higher values of $L_5^r$, while the low value of $F_\eta$ (RQCD21) slightly prefers lower values of $L_5^r$. However, it's quite evident that without additional information no values of $L_5^r$ can be excluded at statistically significant levels. For this reason we need to employ an additional assumption about the values of the LO LECs, either (\ref{Y_eta3pi}) or (\ref{ChQCD_ZY}).

\begin{figure}
    \begin{center}													
    \includegraphics[scale=0.6]{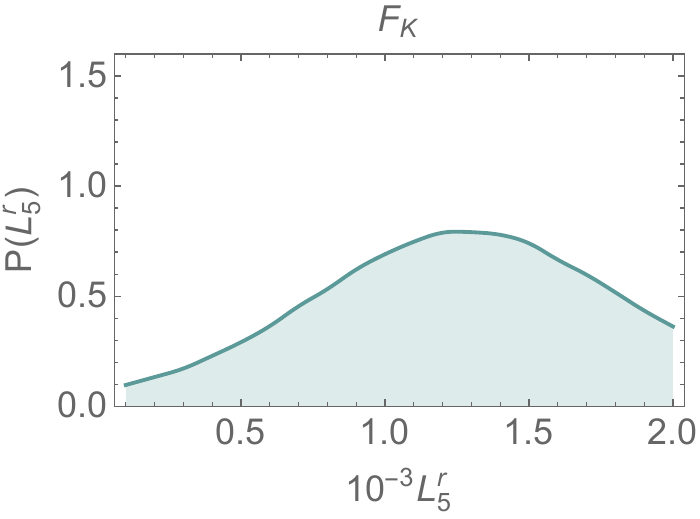}	\\[5pt] 
    \includegraphics[scale=0.6]{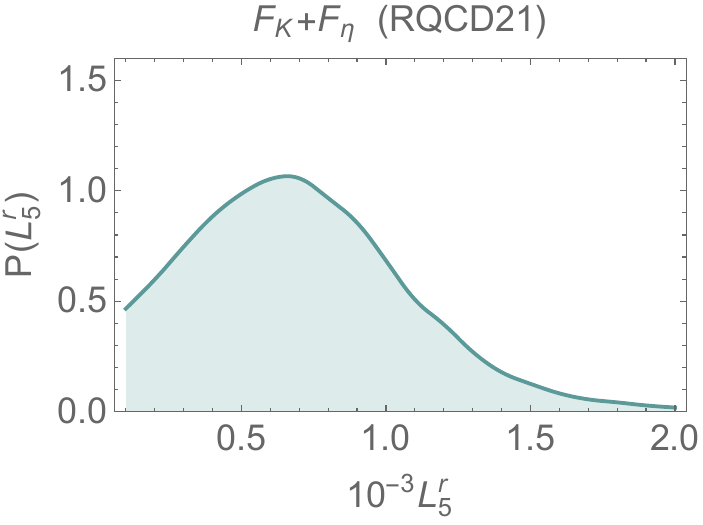}			\end{center}
    \caption{PDFs for $L_5^r$ from $F_K$ and $F_\eta$ for \newline $F_\eta = (1.123 \pm 0.035)F_\pi$ (RQCD21).}
\label{Fig_L5}
\end{figure}

In the first case, using the priors based on the additional input $Y=1.44\pm0.32$ (\ref{Y_eta3pi}) and the relation (\ref{Feta_eq}), depicted in Fig. \ref{Fig_priors_Y_eta3pi} and  Fig. \ref{Fig_delta}, we obtain the following constraints on $L_5^r$ by employing the Bayesian analysis. Figure \ref{Fig_L5} shows our main result, the probability density function for $L_5^r$ using RQCD21 (\ref{RQCD21}), in comparison with only using $F_K$ as an input. Quite clearly, incorporating $F_\eta$ into the analysis has a strong influence.

By approximating with a normal distribution, or alternatively putting a 2$\sigma$ CL bound, we find for all the alternative inputs for $F_\eta$

\begin{align}
    \label{L5(RQCD)}
    L_5^r &= (0.66\pm 0.37)\cdot10^{-3} \qquad \mathrm{(RQCD21)},\\
	L_5^r&<1.34\cdot10^{-3}\ \mathrm{at\ 2\sigma\ CL}, \nonumber\\[10pt]
	L_5^r &= (0.86\pm 0.39)\cdot10^{-3} \qquad \mathrm{(EGMS15)},\\
	L_5^r&<1.60\cdot10^{-3}\ \mathrm{at\ 2\sigma\ CL},\nonumber\\[10pt]
	L_5^r &= (1.43\pm 0.35)\cdot10^{-3} \qquad \mathrm{(EF05)},\\
	L_5^r&>0.78\cdot10^{-3}\ \mathrm{at\ 2\sigma\ CL} \nonumber.
\end{align}

\no In the case of RQCD21 and EGMS15, the obtained values of $L_5^r$  are compatible with both fits BE14/FF14 (\ref{BE14}--\ref{FF14}) and lattice QCD calculations (\ref{HPQCD 13A}--\ref{MILC 10}). However, for a high value of $F_\eta$ from EF05 (\ref{EF05}), we obtain a lower bound for $L_5^r$, which is incompatible with the value from the fit FF14 (\ref{FF14}) -- $L_5^r=(0.5\pm0.07)\cdot10^{-3}$.

Alternatively, using the lattice QCD input for the LO LECs  (\ref{ChQCD_ZY}), depicted on Fig. \ref{Fig_priors_CHQCD} ($\chi$QCD21), we obtain

\begin{align}
	\label{L5(RQCD,ChQCD)}
    L_5^r &= (0.68\pm 0.42)\cdot10^{-3} \qquad \mathrm{(RQCD21,\chi QCD21)},\\
	L_5^r&<1.48\cdot10^{-3}\ \mathrm{at\ 2\sigma\ CL}. \nonumber 
\end{align}

\begin{figure}
    \begin{center}													\includegraphics[scale=0.6]{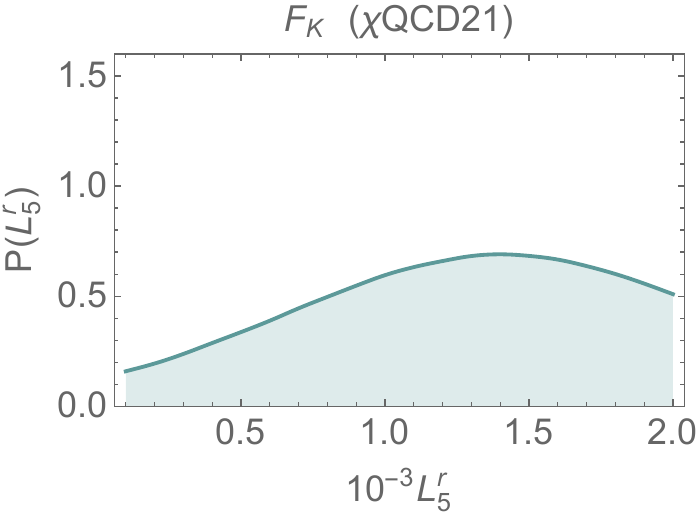} \\[5pt]
    \includegraphics[scale=0.6]{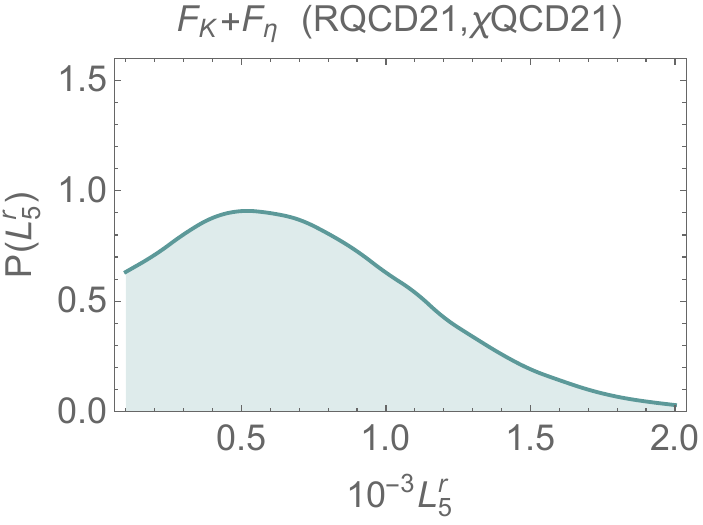}					\end{center}
    \caption{PDFs for $L_5^r$ from $F_K$ and $F_\eta$ (RQCD21), using (\ref{ChQCD_ZY}) ($\chi$QCD21).}
\label{Fig_L5_ChQCD}
\end{figure}

\no The probability distributions are shown in Figure \ref{Fig_L5_ChQCD}. As can be seen, the difference from the previous case is not really significant. It might seem surprising that dramatically restricting $Y$ to a more narrow range (compare Fig. \ref{Fig_priors_CHQCD} vs Fig. \ref{Fig_priors_Y_eta3pi}) actually leads to a slightly larger uncertainty, but that is a result of a weaker dependence of $F_\eta$ on $L_5^r$ at smaller values of $Y$. In other words, a larger value of $Y$ is correlated more strongly with smaller values of $L_5^r$.

\subsection{Extraction of \boldmath{$L_4^r$}}	

\begin{figure}
    \begin{center}
    \includegraphics[scale=0.85]{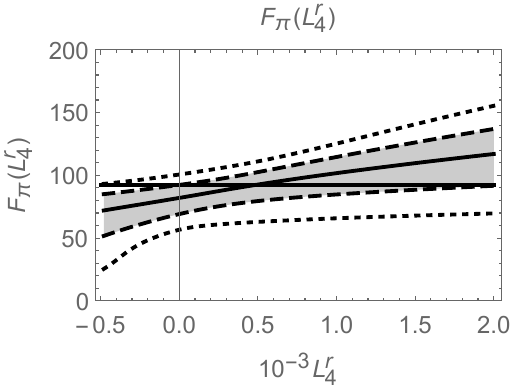} \\[10pt]
    \includegraphics[scale=0.85]{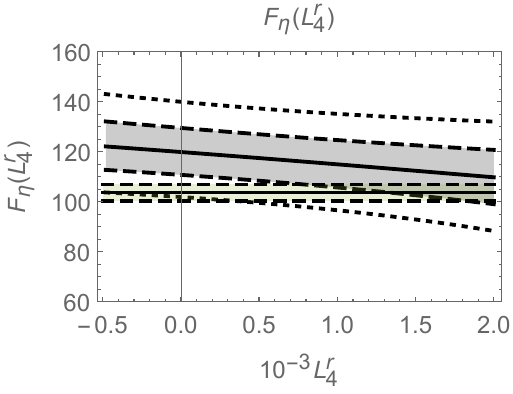}						\end{center}
    \caption{Theoretical predictions for $F_\pi$ and $F_\eta$ with 1$\sigma$ CL (shaded+dashed) and 2$\sigma$ CL (dotted) contours depicted.\\ Horizontal - data from \cite{PDG2020,RQCD:2021qem}.}
\label{Fig_L4_predictions}
\end{figure}

As the last option, we will try to extract information on $L_4^r$. We will essentially repeat the procedure from the last subsection, but in this case, we will use the equation (\ref{FK}) to eliminate $L_5^r$, which gives us a system of two equations for $F_\pi$ and $F_\eta$, the free variables being $Z$, $Y$, $L_4^r(\mu)$ and the remainders $\delta_{F_P}$. Once again, we numerically generated $10^8$ theoretical predictions for $F_\pi$ and $F_\eta$, shown in Figure \ref{Fig_L4_predictions}, using the default priors (\ref{Range_Y}--\ref{P(Y,Z)}) (along with (\ref{delta}--\ref{F_P_data})). As can be seen, while the dependence on $L_4^r$ is markedly different for the two decay constants, our theoretical predictions are compatible with the data in the whole range of values. As in the previous case, we need to employ additional information, i.e. the priors (\ref{Y_eta3pi}) or (\ref{ChQCD_ZY}).

\begin{figure}
    \begin{center}
    \includegraphics[scale=0.6]{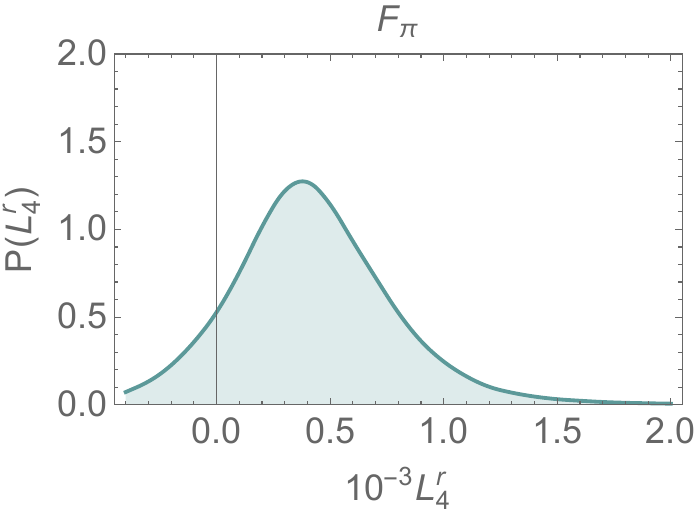} \\[5pt]
    \includegraphics[scale=0.6]{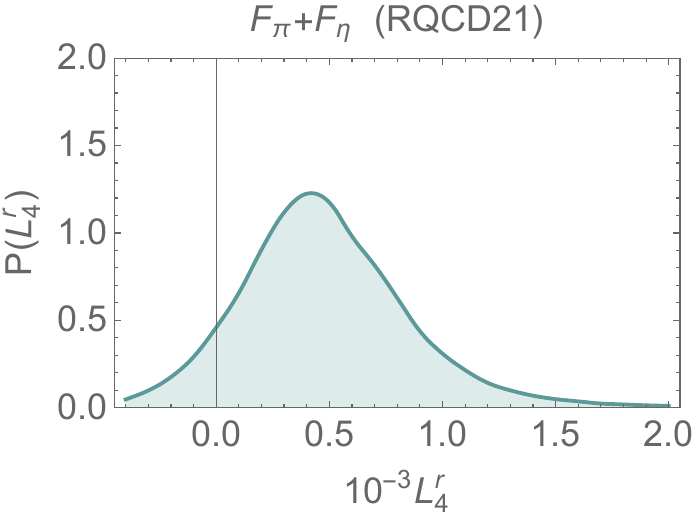}			\end{center}	
    \caption{PDFs for $L_4^r$ from $F_\pi$ and $F_\eta$ for \newline $F_\eta = (1.123 \pm 0.035)F_\pi$ (RQCD21).}
\label{Fig_L4}
\end{figure}

First, using the more conservative choice of priors for the statistical analysis based on (\ref{Y_eta3pi}) (Fig. \ref{Fig_priors_Y_eta3pi}, Fig. \ref{Fig_delta}), we obtained the following probability density functions for $L_4^r$. Figure \ref{Fig_L4} depicts the full result using RQCD21 (\ref{RQCD21}) and also the distribution given solely by $F_\pi$.
\newpage

For all the inputs for $F_\eta$ we get (at $\mu=770$ MeV)

\begin{align}
    L_4^r &= (0.39\pm 0.36)\cdot10^{-3} \qquad (F_\pi\ \mathrm{only)},\\[10pt] 
	\label{L4(RQCD)}
    L_4^r &= (0.44\pm 0.37)\cdot10^{-3} \qquad \mathrm{(RQCD21)},\\[10pt]
	L_4^r &= (0.42\pm 0.36)\cdot10^{-3} \qquad \mathrm{(EGMS15)},\\[10pt]
	L_4^r &= (0.36\pm 0.35)\cdot10^{-3} \qquad \mathrm{(EF05)}.
\end{align}	

\no Interestingly, in this case the strongest constraint is generated by the chiral expansion of $F_\pi$ (\ref{Fpi}) and adding $F_\eta$ into the analysis does not make a marked change. As can be seen, varying the input for $F_\eta$ does not have a significant impact and all results are compatible with both the fits BE14/FF14 (\ref{BE14}--\ref{FF14}) and lattice QCD calculations (\ref{HPQCD 13A}--\ref{MILC 10}).

Using the alternative priors for the leading order LECs from $\chi$QCD21 (\ref{ChQCD_ZY}) (Fig. \ref{Fig_priors_CHQCD}), we obtain

\begin{align}
    L_4^r &= (0.38\pm 0.25)\cdot10^{-3} \qquad (F_\pi\ \mathrm{only,\chi QCD21)},\\[10pt]
    \label{L4(RQCD,ChQCD)}
    L_4^r &= (0.46\pm 0.24)\cdot10^{-3} \qquad \mathrm{(RQCD21,\chi QCD21)}.
\end{align}

\begin{figure}
    \begin{center}
    \includegraphics[scale=0.6]{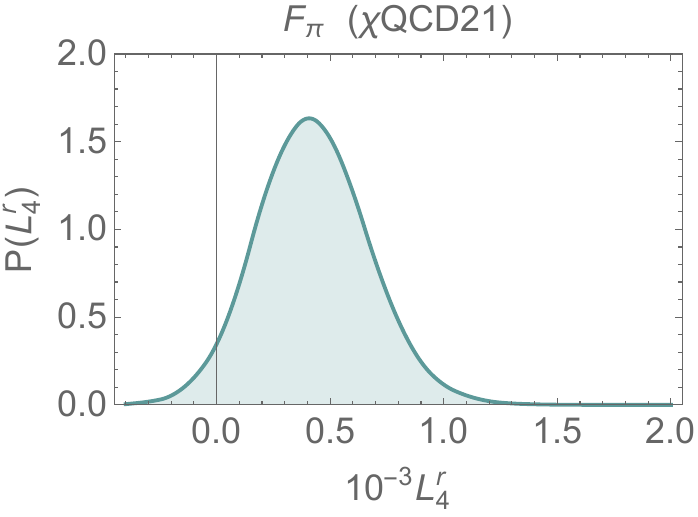}	\\[5pt]
    \includegraphics[scale=0.6]{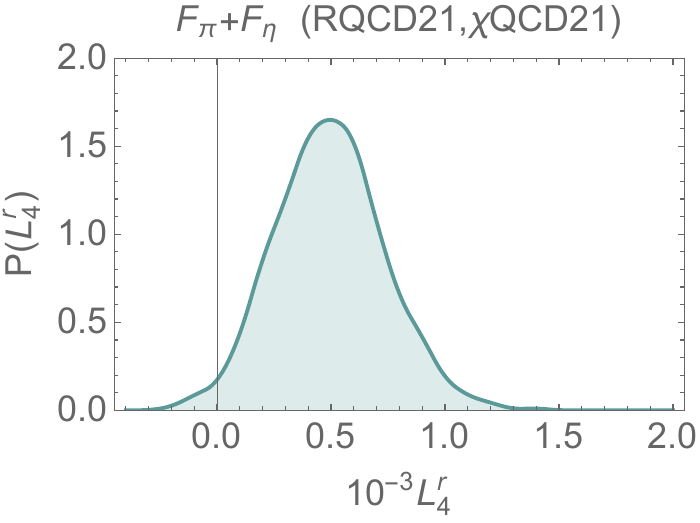}
    \end{center}	
    \caption{PDFs for $L_4^r$ from $F_\pi$ and $F_\eta$ (RQCD21), using (\ref{ChQCD_ZY}) ($\chi$QCD21).}
\label{Fig_L4_ChQCD}
\end{figure}

\no The PDFs can be found in Figure \ref{Fig_L4_ChQCD}. In contrast to $L_5^r$, in this case the more restricted range for the LO LECs leads to somewhat smaller error bars.  

One might naturally ask whether the obtained results for $L_4^r$ and $L_5^r$ also signify an update on the priors of $Z$ and $Y$ and thus could shed some light on the pattern of chiral symmetry breaking at the leading order. We have investigated this possibility, but the updated PDFs (not shown here) are not significantly different from the priors. This is not surprising, given that the results for the NLO LEC's do not strongly depend on the alternative choice of the priors ((\ref{Y_eta3pi}) or (\ref{ChQCD_ZY})), as can be seen from comparing (\ref{L5(RQCD)}) vs (\ref{L5(RQCD,ChQCD)}) and (\ref{L4(RQCD)}) vs (\ref{L4(RQCD,ChQCD)}). The largest effect comes from excluding very low values of $Y$, which both cases do. In other words, the rest of the uncertainties, mainly coming from the higher order remainders, are large enough that the chiral order parameters can't be constrained purely from inputs for the decay constants without additional information about the higher orders.

Finally, it is also possible to illustrate the correlation naturally expected between $L_4^r$ and $Z$ from (\ref{Fpi}--\ref{Feta}). When setting $Z$ by hand and restricting $Y$ from below, we have obtained the following limits:

\begin{align}
\label{L4-Z corr}
    L_4^r&<0.38\cdot10^{-3}\ \mathrm{at\ 2\sigma\ CL}\\
    \nonumber &\hsp{1.5cm}\mathrm{(RQCD21, Y>0.8, Z=0.8)},\\[10pt]
    L_4^r&<0.88\cdot10^{-3}\ \mathrm{at\ 2\sigma\ CL}\\
    \nonumber &\hsp{1.5cm}\mathrm{(RQCD21, Y>0.8, Z=0.5).}
\end{align}

\no The first scenario essentially corresponds to the limits provided by the $\eta\to 3\pi$ decays \cite{Kolesar:2017xrl}, discussed above. As can be seen from (\ref{L4-Z corr}), such a high value of $Z$ ($F_0\approx82$ MeV) restricts $L_4^r$ much more strongly and would be in fact incompatible with a high value $L_4^r = (0.76\pm 0.18)\cdot10^{-3}$, obtained by the fit FF14 (\ref{FF14}). While this result might not be surprising, we are able demonstrate it quantitatively, with taking all the uncertainties into account.

\section{Summary \label{sec:Summary}}

We have investigated the sector of decay constants of the octet of light pseudoscalar mesons in the framework of 'resummed' chiral perturbation theory. Our theoretical prediction for the $SU(3)$ decay constant of the $\eta$ meson is 

\begin{equation}
	F_\eta = 117.5\pm 9.4\ \mathrm{MeV} = (1.28\pm0.10)F_\pi,
\end{equation}

\no which is compatible with recent determinations \cite{Escribano:2005qq,Escribano:2015yup,RQCD:2021qem}.

Utilizing these determinations as inputs for $F_\eta$, we have applied Bayesian statistical inference to extract the values of next-to-leading order low-energy constants $L_4^r$, $L_5^r$ and higher order remainders $\delta_{F_K}$ and $\delta_{F_\eta}$. $L_5^r$ was assumed to be positive, while $L_4^r>{L_4^r}^{(crit)}=-0.5\times 10^{-3}$. 

By using the most recent lattice QCD data from the RQCD Collaboration \cite{RQCD:2021qem}, which provided us with the best estimate $F_\eta = (1.123 \pm 0.035)F_\pi$, we have obtained our main result (at $\mu=770$ MeV):

\begin{align}
    L_4^r &= (0.44\pm 0.37)\cdot10^{-3} \qquad \mathrm{(RQCD21)},
    \\[10pt]
    L_5^r &= (0.66\pm 0.37)\cdot10^{-3} \qquad \mathrm{(RQCD21)}, \\ 
    L_5^r&<1.34\cdot10^{-3}\ \mathrm{at\ 2\sigma\ CL}. \nonumber
\end{align}
\begin{align}
    \delta_{F_K} &= 0.10 \pm 0.07, \nonumber\\
    \delta_{F_\eta} &= -0.08 \pm 0.08 \qquad\qquad \mathrm{(RQCD21)},\\
    \rho &= 0.71. \nonumber
\end{align}

\no These results have used conservative estimates for the priors of the low-energy constants at the leading order ($F_0$ and $B_0$).

Alternatively, we have used a recent computation of the leading order LECs $F_0$ and $B_0$ by the $\chi$QCD Collaboration \cite{CHQCD:2021pql} as an additional input. Though the work has not been fully published yet, it has been cited by the Flavour Lattice Averaging Group \cite{FLAG:2021npn}. In this case we have obtained:

\begin{align}
    L_4^r &= (0.46\pm 0.24)\cdot10^{-3} \qquad \mathrm{(RQCD21,\chi QCD21)},\\[10pt]
    L_5^r &= (0.68\pm 0.42)\cdot10^{-3} \qquad \mathrm{(RQCD21,\chi QCD21)}, \\  
    L_5^r&<1.48\cdot10^{-3}\ \mathrm{at\ 2\sigma\ CL}. \nonumber
\end{align}

As our main conclusion, all these values are compatible within uncertainties with the most recent standard $\chi$PT fits BE14 and FF14 \cite{Bijnens:2014lea}, as well as the lattice QCD computations cited by the FLAG review \cite{FLAG:2021npn}. So quite clearly, while we independently confirm the generic range of values available in the literature, an additional source of information needs to be found in order to pin down the values of the low energy constants more precisely.

However, when testing inputs for $F_\eta$ from phenomenology, we have found some tension if a high value of $F_\eta = (1.38 \pm 0.05)F_\pi$ (EF05) \cite{Escribano:2005qq} was assumed. This lead to a negative sign of the remainder $\delta_{F_K}$, while both fits BE14 and FF14 in \cite{Bijnens:2014lea} have positive NNLO contributions for $F_K$. In addition, such a high value of $F_\eta$ produced a lower bound $L_5^r<0.78\cdot10^{-3}\ (\mathrm{at\ 2\sigma\ CL}$), which is incompatible with the value from the fit FF14 ($L_5^r=(0.5\pm0.07)\cdot10^{-3}$). \\

\no\textbf{Acknowledgement:} We would like to thank K.~Kampf and J.~Novotn\'y for their time and valuable inputs.\\

\no\textbf{Data Availability Statement:} The statistical datasets used in the work were numerically generated by a Mathematica script, according to the procedures and inputs described in the text. We do not store the raw data, but the script can be shared upon request.

\end{document}